\documentclass[shortnote,twocolumn]{jpsj2} 
\usepackage{color}
\title{
Projection Method for Exact Diagonalization 
in the Strong Coupling Limit for 
Frustrated Models with Highly Degenerate Ground States 
}

\author{
Keisuke \textsc{Matsuda}$^{1}$,
Shin \textsc{Miyahara}$^{2}$,
and Nobuo 
\textsc{Furukawa}$^{1,2}$
}

\inst{
$^{1}$Department of Physics and Mathematics, Aoyama Gakuin University,
Sagamihara, Kanagawa 229-8558 \\
$^{2}$Multiferroics Project (MF), ERATO,
Japan Science and Technology Agency (JST), 
c/o Department of Applied Physics, The University of Tokyo, 
Tokyo 113-8656
}

\abst{
}

\kword{
geometrical frustration, projection, exact diagonalization,
$t$-$V$ model, XXZ Heisenberg model
}

\begin{document}
\maketitle

Numerical calculations play an important role
in clarifying the features in strongly correlated electron 
systems.~\cite{alet05} 
In particular, quantum Monte Carlo (QMC) and
density matrix renormalization group (DMRG) methods
are often adopted to quantum spin and electron systems.
However, most of them are useful only for the 
system without geometrical frustration or in low dimension.

Exact diagonalization (ED) is applicable even in
two- or three-dimensional frustrated systems,
although there exist limitations to system sizes,
{\it e.g.}, $36$ sites in two-dimensional spin-1/2 kagom\'{e} 
antiferromagnetic Heisenberg systems at present.
To increase the system size, 
calculations with some approximation have been attempted.
One of the successful methods is a numerical 
contractor renormalization 
method~\cite{capponi02,capponi04}. 
In this method, an effective Hamiltonian is constructed on
a truncated local basis set with ED and the 
calculated system size increases up to 
$48$ sites in a spin-1/2 kagom\'{e} system~\cite{capponi04}.  

\begin{figure}
 \begin{center}
  \includegraphics[width=0.5\linewidth,keepaspectratio,clip]{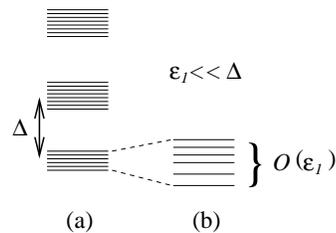}
 \end{center}
 \caption{ (a) Macroscopic degenerate ground state
   with excitation energy $\Delta$.
   (b) Calculated energy within the projected 
   subspace for $\varepsilon_1 \ll \Delta$.}
 \label{fig:sketch}
\end{figure}

In this paper, we carry out a new technique named
projection ED and indicate that it is applicable
to the frustrated models in the strong coupling limit
with highly degenerate ground states in a general way.  
Let us consider the Hamiltonian 
\begin{align} 
 \mathcal{H}
 =& 
 \mathcal{H}_0 + \mathcal{H}_1.
 \label{eq:hmlt}
\end{align}
Here, the Hamiltonian $\mathcal{H}_0$ has classical degrees of freedom
and its eigenvalues are discrete. In addition, 
its ground state is assumed to be macroscopically degenerate and
to have an excitation gap $\Delta$
(see Fig.~\ref{fig:sketch}(a)).
$\mathcal{H}_1$ is a quantum term 
that hybridizes the eigenstates of $\mathcal{H}_0$,
which will be treated perturbatively.

As typical examples,
we consider
frustrated XXZ spin-1/2 Heisenberg and spinless fermion $t$-$V$ models
in a strong coupling region, {\em i.e.},
$|J_{xy}| \ll J_{z} $ in the former model 
and $|t| \ll V$ in the latter model,
where $J_{xy}$ and $t$ give 
perturbation terms ${\mathcal H}_1$.
The term ${\mathcal H}_0$ corresponds to an Ising model. 
On some geometrically frustrated lattices, {\em i.e.}, 
on triangular, kagom\'{e} and pyrochlore lattices,
the models with ${\mathcal H}_1 = 0$
have macroscopically degenerate ground states
when interactions are antiferromagnetic, {\em i.e.},
$J_{z} >0$ or $V>0$.\cite{wannier50,syozi51,kano53,anderson56}

For $\Delta \gg \varepsilon_1$,
where $\varepsilon_1$ is the energy scale of ${\mathcal H}_1$,
we may apply the perturbation theory to describe
low energy states as those spanned within the Hilbert space projected to
the degenerate ground states of $\mathcal{H}_0$,
as shown in Fig.~\ref{fig:sketch}(b). 
Within the second order of the perturbation of 
${\mathcal H_1}$, we can obtain an effective Hamiltonian:
\begin{align}
 \tilde{\mathcal{H}} = &
 E_0 + 
 \mathcal{P} 
 \mathcal{H}_1
 \mathcal{P} 
 -
 \mathcal{P} \mathcal{H}_1 \mathcal{Q} 
 \frac{1}{\mathcal{H}_0}
 \mathcal{Q} \mathcal{H}_1 \mathcal{P}, 
 \label{eq:effective_hmlt}
\end{align}
where $\mathcal{P}$ is a projection operator,
which takes $1$ in case that the applied state is 
one of the ground states of the Ising model, 
and $0$ for higher energy states.
$\mathcal{Q}$ is the complement of $\mathcal{P}$ and 
satisfies $\mathcal{P} + \mathcal{Q} = 1$.
$E_0$ is the ground state energy of 
the classical model $\mathcal{H}_0$.

In these cases, we have difficulties caused by 
(i) macroscopic degeneracy in the ground state and
(ii) the nonlocal property of $ \mathcal{P} $.
Due to the former reason,
analytical approaches for perturbative
calculations are extremely difficult. Therefore, we need to
resort to numerical calculations.
To the advantage of ED,
the number of ground states of $\mathcal{H}_0$ is much smaller than 
the total dimension of the original Hilbert space of 
${\cal H}$, in general.
Thus, it becomes possible to handle a much larger system than 
in the case of conventional ED.

The latter difficulty originates from the fact that,
unlike the simple cases such as the Hubbard 
model at $U \rightarrow \infty$,
the unperturbed ground states themselves are highly nontrivial.
This gives
complex forms for analytical expressions of
the projection operators $\mathcal{P}$ and
$\mathcal{Q}$, and hence, the overall 
perturbation terms.
In addition,  since the ground states depends 
on lattice structures,
the formula also depends on them
and is thus nonuniversal.

To avoid such difficulties, we introduce ED,
which consequently makes it possible 
to treat the problem in general.
Hereafter, we show the details based on the calculations
for the spinless fermion $t$-$V$ model 
on a triangular lattice as an example.
A Hamiltonian 
can be written as
\begin{align} 
 \mathcal{H}_0
 = &
 V \sum_{\langle ij \rangle} 
 \left( n_i -\tfrac{1}{2} \right) \left( n_j -\tfrac{1}{2} \right),
 \label{eq:hmlt_0}
 \\
 \mathcal{H}_1
 = &
 -t \sum_{\langle ij \rangle} 
 \left( c^\dagger_i c_j +{\mathrm H.c.} \right). 
 \label{eq:hmlt_1}
\end{align}
Here, $c_{i}^{\dagger}$ ($c_{i}$) is the creation (annihilation) operator 
of fermions at site $i$ and $n_{i}=c^{\dagger}_{i}c_{i}$. 
$V$ is a repulsive interaction between nearest neighboring sites.
We allow hopping between nearest neighboring sites
on a triangular lattice  
with an amplitude $t>0$.

Lanczos procedure has been performed by applying 
the Hamiltonian ${\cal H}_1$, 
which is an $\mathcal{O}(M)$ calculation.
Here, $M$ is the number of ground states of $\mathcal{H}_0$
at a given fermion density.
For the hopping from site $i$ to $j$, 
a locally modified energy can be obtained
by confirming the 10-site configuration 
including 
8 neighboring sites of $i$ and/or $j$ on a triangular lattice.
When the modified state belongs to the ground state of ${\cal H}_0$, 
the hopping term contributes as the first-order perturbation
term in eq.~(\ref{eq:effective_hmlt}).
In the case that the modified state is an excited state of ${\cal H}_0$,
further hopping processes 
which return the excited state to one of the ground states,
lead to the second-order perturbation term in eq.~(\ref{eq:effective_hmlt}).
Generally, the possibility of such a correlated hopping 
can be confirmed by considering the 13-site configuration at most.
Since what we need is the energy of a locally modified
state for the Ising model,
the type of calculation is an $\mathcal{O}(1)$ calculation.
In this manner, projection ED
is generally an $\mathcal{O}(M)$ calculation.
Note that, even if we got the formula in eq.~(\ref{eq:effective_hmlt}) 
analytically, its direct Lanczos operation is also
$\mathcal{O}(M)$.

\begin{figure}
 \begin{center}
  \includegraphics[width=0.7\linewidth,keepaspectratio,clip]{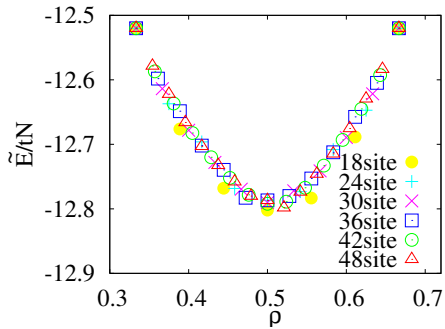}
 \end{center}
 \caption{
  (Color online) Energies $\tilde{E}/tN$ at density $\rho$ on triangular 
  spinless fermion $t$-$V$ model for $V/t = 50$.
  The results for up to 48-site clusters are shown.
  }
 \label{fig:mu-N}
\end{figure}

As shown in Table~\ref{tbl:state_number},
the number of states $M$ 
is much smaller than that of conventional ED.
Thus, we can reach a much larger cluster than in the case of 
conventional ED.
For example, we could carry out the calculation up to 48-site clusters
in the second order of perturbation easily.
Energies $\tilde{E}/t$ as a function of 
fermion densities are shown in Fig.~\ref{fig:mu-N}.
Here, the calculations have been performed for the numbers of 
sites $N = 18, 24, 30, 36, 42$ and $48$
at the densities $1/3 \leq \rho \leq 2/3$ for $V/t = 50$.
Note that we can handle a much larger system 
on kagom\'{e} and pyrochlore lattices
by also applying projection ED.

\begin{table}[tb]
 \begin{tabular}{|r|r|r|}
  \hline
  \multicolumn{1}{|c|}{$N$}  &    
  \multicolumn{1}{c|}{$M$ (ED)} & 
  \multicolumn{1}{c|}{$M$ (projection ED)} \\ \hline\hline
  12       &                924 &                      66 \\ \hline
  18       &             48,620 &                     386 \\ \hline
  24       &          2,704,156 &                   2,142 \\ \hline
  30       &        155,117,520 &                  13,694 \\ \hline
  36       &      9,075,135,300 &                  84,486 \\ \hline
  42       &    538,257,874,440 &                 553,574 \\ \hline
  48       & 32,247,603,683,100 &               3,720,042 \\ \hline
 \end{tabular}
 \caption{
   Number of states at half-filling for conventional 
   and projection EDs on N-site clusters 
   for the triangular $t$-$V$ model.
 }
 \label{tbl:state_number}
\end{table}

\begin{figure}[tb]
 \begin{center}
  \includegraphics[width=0.78\linewidth,keepaspectratio,clip]{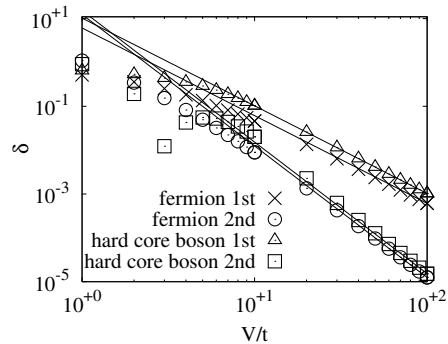}
 \end{center}
 \caption{
 Comparison between conventional and projection EDs on 
 the spinless fermion $t$-$V$ and hard core boson models  
 on the triangular lattice. The calculation has been performed
 at half-filling for both first- and second-order perturbations.
 For $V/t \gtrsim 10$, the energy differences 
 are proportional to $(t/V)^2$ for the first-order calculation and  
 $(t/V)^3$ for the second-order calculation.
 The fitting results are shown by the lines.
 }
 \label{fig:comparision}
\end{figure}

The accuracy of our method can be confirmed
by comparison with conventional ED in small clusters.
We calculate the energy differences 
between projection and conventional EDs using 
$\delta = | ( \tilde{E}  - E  )/ E |$, 
where $\tilde{E}$ is the energy calculated by projection ED,
and $E$ is the energy obtained by ED. The results for a 24-site cluster 
are shown in Fig.~\ref{fig:comparision}. 
In the parameter region, where
the $n$-th order of perturbation is 
sufficiently accurate, $\tilde{E}$ is consistent with $E$ 
within $\mathcal{O}((t/V)^n)$, {\em i.e.},
the errors $\delta$ are proportional to $(t/V)^{n+1}$.
As shown in the figure, we can obtain a good agreement 
in the second-order perturbation for $V/t \gtrsim 10$,
where the errors are proportional to  $(t/V)^3$. 
In the figure,
the results for a hard core boson 
model on a triangular lattice are also shown. 

Finally, we mention that researches of the hard core boson 
and $t$-$V$ models on the triangular lattice 
by projection ED are significant.
These models have been paid much attention to,
after the prediction of the realization of a new type of 
ground state, {\em i.e.}, a supersolid state 
in the hard core boson model~\cite{wessel05,heidarian05,melko05}
and a pin-ball liquid state in the $t$-$V$ model~\cite{hotta06}. 
In the former, most of the calculations have been performed by QMC simulation,
and in the latter, ED has been performed. In QMC, it is difficult to
calculate in parameter ranges $V/t \gtrsim 10$. For ED, the system size 
is limited. Thus, our proposal should be important to clarifying 
the features at certain parameter ranges or system sizes, 
at which previous methods cannot be applied.

We would like to thank M. Miyazaki and C. Hotta
for stimulating discussions.

\end{document}